\documentclass[12pt]{article}
\usepackage{graphicx}
\usepackage{cite}

\def \bea{\begin{eqnarray}}
\def \beq{\begin{equation}}

\def \eea{\end{eqnarray}}
\def \eeq{\end{equation}}

\def \({\left(}
\def \){\right)}
\def \[{\left[}
\def \]{\right]}

\def \bea{\begin{eqnarray}}
\def \beq{\begin{equation}}
\def \eea{\end{eqnarray}}
\def \eeq{\end{equation}}

\begin{document}
\rightline{EFI 14-12}
\rightline{July 2014}
\rightline{arXiv:1405.2885}
\smallskip
\centerline{\bf ASYMMETRY IN $\Lambda_b$ AND $\bar \Lambda_b$ PRODUCTION}
\bigskip
\centerline{Jonathan L. Rosner}
\centerline{\it Enrico Fermi Institute and Department of Physics}
\centerline{\it University of Chicago, 5620 S. Ellis Avenue, Chicago, IL 60637}
\bigskip
\begin{quote}
In CMS data at the CERN Large Hadron Collider, the ratio $\sigma(pp \to \bar
\Lambda_b X)/\sigma(p p \to \Lambda_b X)$ appears to fall as the baryons become
more forward.  Mechanisms which could give rise to this effect are discussed.
It is urged that the same physics be explored in data from the ATLAS and LHCb
Detectors at CERN and the Fermilab Tevatron proton-antiproton collider.  In
the latter, if such leading-baryon effects are present, one expects $\Lambda_b$
to be preferentially produced in the direction of the proton and $\bar
\Lambda_b$ to be preferentially produced in the direction of the antiproton.
\end{quote}

\leftline{PACS numbers: 14.20.Mr, 14.65.Fy, 12.38.Aw, 12.38.Qk}
\bigskip

\section{Introduction}

The production of heavy baryons and antibaryons in hadronic collisions
has posed a theoretical puzzle for a number of years, ever since the
observation at the CERN Intersecting Storage Rings (ISR) of the charmed
baryon $\Lambda_c$ \cite{Lockman:1979aj,Chauvat:1987kb}.  Lowest-order QCD
involving the subprocesses $q \bar q \to c \bar c$ and $g g \to c \bar c$,
where $q$ is a light quark ($u,d,s$) and $g$ is a gluon, would predict equal
cross sections for $\Lambda_c$ and $\bar \Lambda_c$ for each value of $x_F$
and $p_T$.  However, production of $\Lambda_c$ in proton-proton collisions at
the ISR is favored over that of $\bar \Lambda_c$, indicating the presence of
 non-perturbative final-state interactions such as those occurring in a QCD
string model like PYTHIA \cite{Bengtsson:1987kr,Sjostrand:2006za}.  (For an
early overview of fragmentation models see \cite{Sjostrand:1987xj}.)

Asymmetries in the production of bottom quarks at the LHC were investigated
some time ago \cite{Norrbin:2000jy} and found to be negligible except in the
very forward direction (beyond the reach of LHCb).  Methods employed were
the Lund string fragmentation model \cite{Andersson:1983ia} and the
intrinsic heavy quark model \cite{Brodsky:1980pb,Brodsky:1981se}.

At $\sqrt{s} = 7$ TeV the LHCb Collaboration \cite{Aaij:2012cy} finds a
production asymmetry
$A_P = [\sigma(D_s^+)-\sigma(D_s^-)]/[\sigma(D_s^+)+\sigma(D_s^-)] =
(-0.33\pm 0.22 \pm 0.10)\%$
for $2.0 \le y \le 4.5$, exhibiting no preference for a leading-quark effect.
Recently the production of $\Lambda_b$ and $\bar \Lambda_b$ has been studied
by the CMS Collaboration at the CERN Large Hadron Collider (LHC).  While
no significant difference between $\Lambda_b$ and $\bar \Lambda_b$ production
is seen in the central region with $|y^{\Lambda_b}| \le 1.5$
\cite{Chatrchyan:2012xg}, the $\bar \Lambda_b$ is produced only about 2/3 as
frequently as the $\Lambda_b$ in the most forward rapidity bin $1.5 \le
|y^{\Lambda_b}| \le 2.0$.  The present note calls attention to a simple way of
evaluating the string-based fragmentation mechanism leading to an asymmetry,
and to urge that this asymmetry be examined in the data of ATLAS and LHCb at
the LHC and CDF and D0 at the Fermilab Tevatron.

In Section II we review recent data on $\Lambda_b$ and $\bar \Lambda_b$
production at the LHC.  We then recall in Section III a ``color reconnection''
mechanism proposed recently \cite{Rosner:2012pi} in the context of a
forward-backward asymmetry in top quark production at the Tevatron observed by
CDF \cite{Aaltonen:2011kc,Aaltonen:2012it,CDF:2013gna,Aaltonen:2013vaf} and D0
\cite{Abazov:2011rq,Abazov:2012oxa,Abazov:2013wxa,Abazov:2014vga,%
Abazov:2014oea,Abazov:2014cca}.  Effects of this mechanism should be
contained in any model which seeks to predict the production of $\Lambda_b$
and $\bar \Lambda_b$ at hadron colliders.  Questions of $p_T$ and $y$
dependence, and possible polarization effects, are discussed very briefly in
Section IV.  We close in Section V by urging such studies at ATLAS, LHCb, and
the Tevatron.

\section{Recent data}

\begin{figure}
\begin{center}
\includegraphics[width=0.6\textwidth]{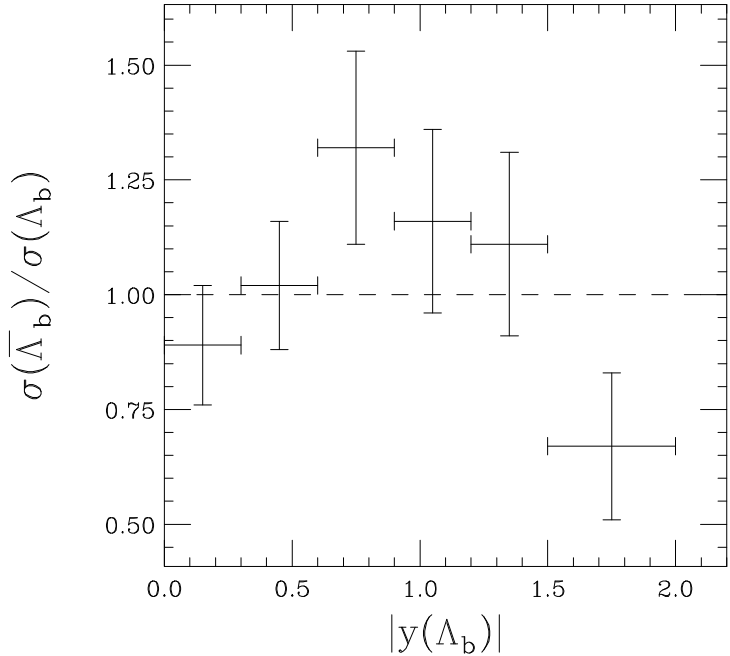}
\end{center}
\caption{Ratio of $\bar \Lambda_b$ and $\Lambda_b$ cross sections reported
by the CMS Collaboration \cite{Chatrchyan:2012xg} as a function of rapidity.
Only statistical errors are shown; systematic errors for the points are
$\pm 0.09,\pm0.09,\pm0.13,\pm0.12,\pm0.15,\pm0.16$, respectively.  The dashed
line denotes a ratio of 1.
\label{fig:cmsrat}}
\end{figure}

The production of $\Lambda_b$ and $\bar \Lambda_b$ has been studied at the
LHC by the CMS Collaboration \cite{Chatrchyan:2012xg}, based on an integrated
luminosity of 1.96 fb$^{-1}$ at $\sqrt{s} = 7$ TeV.  The reported ratio of
$\bar \Lambda_b$ and $\Lambda_b$ cross sections
is illustrated in Fig.\ \ref{fig:cmsrat} as a function of $|y(\Lambda_b)|$,
the $\Lambda_b$ rapidity.  Although no significant variation with
$|y(\Lambda_b)|$ is claimed in Ref.\ \cite{Chatrchyan:2012xg}, one can also
see a modest decrease in the ratio in the most forward rapidity bin, as
pointed out in Ref.\ \cite{Rosner:2012pi}.

It was noted in Ref.\ \cite{Rosner:2012pi} that the LHCb Collaboration was in
an ideal position to extend this measurement to larger $|y|$, where a string
fragmentation picture would predict a growing predominance of $\Lambda_b$
over $\bar \Lambda_b$.  If the trend suggested by CMS contiunes to higher $y$,
the $\bar \Lambda_b$ cross section at LHCb would be no more than 2/3 that of
the $\Lambda_b$, suggesting that different production mechanisms were at
work in the central and forward directions.  Some possibilities for these
mechanisms are described in the next Section.  It is notable that the
decreased ratio of cross sections suggested by the CMS data is not reproduced
by the POWHEG or PYTHIA Monte Carlo predictions.

We note briefly some other LHC $\Lambda_b$ data in $\bar p p$ collisions at
$\sqrt{s} = 7$ TeV, to be discussed in more detail in Section IV.  The CMS
Collaboration has studied the polarization of $\Lambda_b$ and $\bar \Lambda_b$
with a sample corresponding to 5.1 fb$^{-1}$ of integrated luminosity
\cite{Rikova:2013}.  The ATLAS Collaboration \cite{Aad:2014iba} has published a
study of $\Lambda_b$ and $\bar \Lambda_b$ polarization with 4.6 fb$^{-1}$ of
data, but without stating the relative production fractions of $\Lambda_b$
and $\bar \Lambda_b$.  Finally, the LHCb Collaboration \cite{Aaij:2013oxa}
has studied $\Lambda_b$ and $\bar \Lambda_b$ polarization with 1 fb$^{-1}$. 

\section{Production mechanisms}

\subsection{Mechanisms without asymmetry}

The subprocesses $q \bar q \to b \bar b$ and $g g \to b \bar b$, followed by
fragmentation of a $b$ quark into $\Lambda_b$ or a $\bar b$ quark into
$\bar \Lambda_b$, do not lead to an asymmetry between baryon and antibaryon
production.  One might expect these processes to dominate in production of
heavy baryons with small $|y|$ and large $p_T$.  Some additional processes are
contributing to $\Lambda_b$ production at small $p_T$; its cross section falls
off more rapidly with increasing $p_T$ than the cross sections for $B$-flavored
mesons \cite{Chatrchyan:2012xg}.

\subsection{Quasi-diffractive excitation}

\begin{figure}
\begin{center}
\includegraphics[width=0.5\textwidth]{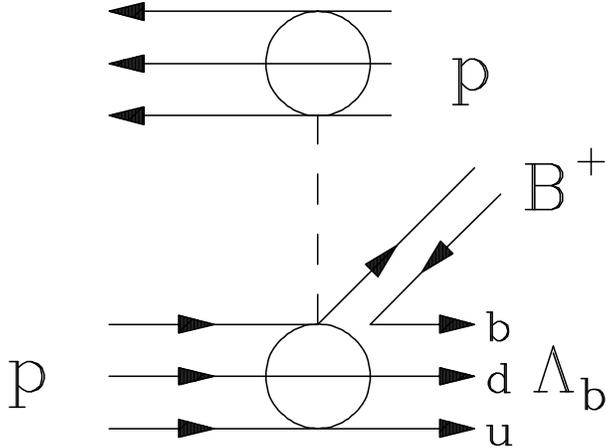}
\end{center}
\caption{Quasi-diffractive production of $\Lambda_b$ along the direction
of a proton beam.  The circles denote vertices for exchange of a Pomeron,
denoted by the dashed line.
\label{fig:diffrac}}
\end{figure}

In Fig.\ \ref{fig:diffrac} we illustrate a mechanism which may be expected
to contribute to forward heavy baryon production and will favor production
of $\Lambda_b$ by protons and $\bar \Lambda_b$ by antiprotons.  The figure
suggests that a forward $\Lambda_b$ often will be accompanied by a forward
$B^+$ or the decay products of an excited $B^+$.  This mechanism has some
features in common with the intrinsic heavy quark model \cite{Brodsky:1980pb,%
Brodsky:1981se}, in the sense that a heavy forward baryon is more likely to
contain a $b$ quark rather than a $\bar b$.  

\subsection{Interaction with spectator quark}

\begin{figure}
\begin{center}
\includegraphics[width=0.6\textwidth]{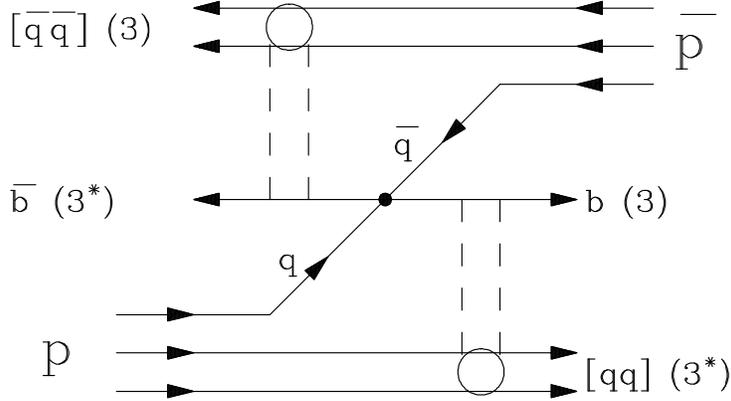}
\end{center}
\caption{Interaction of final-state heavy quark with spectator system,
as proposed in Ref.\ \cite{Rosner:2012pi}.  The pairs of dashed lines denote
QCD strings connecting the final-state heavy quarks to the spectator systems.
\label{fig:dragfig}}
\end{figure}

The final-state interaction of the heavy quark with the proton remnants
\cite{Norrbin:2000jy,Andersson:1983ia} was noted in the case of $t \bar t$
production in Ref.\ \cite{Rosner:2012pi}.  (See also \cite{Brodsky:2012rw}.)
This mechanism is illustrated in Fig.\ \ref{fig:dragfig}.  The effect of this
process on the apparent asymmetry in $t \bar t$ production at the Tevatron was
seen to be too small to account for the asymmetry claimed initially by both
Collider collaborations, but in a recent report by D0 the asymmetry no longer
conflicts with the standard model \cite{Abazov:2014cca}.  

We retrace the argument presented in Ref.\ \cite{Rosner:2012pi} for the
``drag'' exerted by a QCD string on a heavy quark produced through the
process illustrated in Fig.\ \ref{fig:dragfig}.  We first calculate in the
frame where the longitudinal momentum of the heavy quark is zero.  A result
expressed in terms of rapidity then is invariant under boosts along the $z$
axis.

\begin{figure}
\begin{center}
\includegraphics[width=0.6\textwidth]{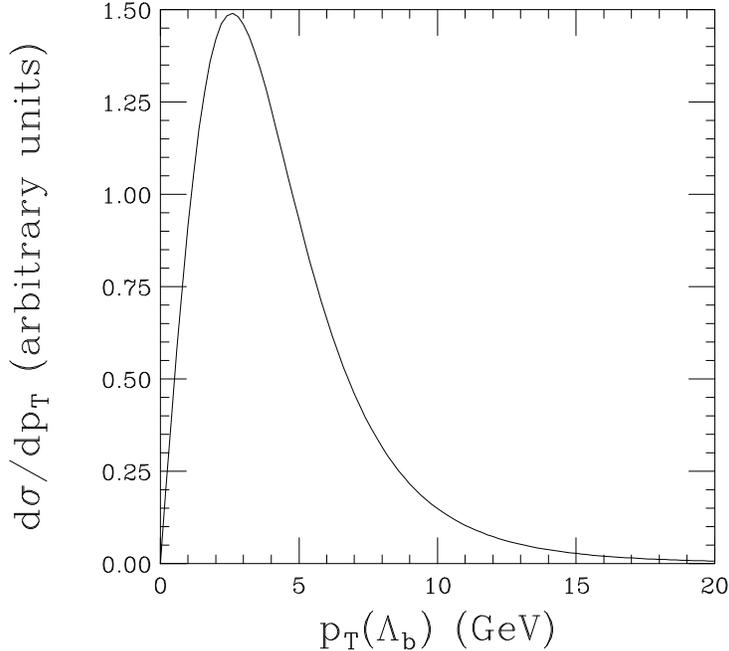}
\end{center}
\caption{Distribution (``Tsallis function'' \cite{Tsallis}) fitting
$p_T$-dependence of $\Lambda_b$ production reported by the CMS
Collaboration \cite{Chatrchyan:2012xg} at $\sqrt{s} = 7$ TeV.
\label{fig:ptdist}}
\end{figure}

A QCD string breaks when it reaches a length of about 1.5 fm
\cite{Rosner:1996xz}.  If its end attached to the remnant travels with respect
to the other end at the speed of light, it acts for a time
\beq
t=\frac{1.5 \times 10^{-15}~{\rm m}}{3 \times 10^8~({\rm m/s})} = 5 \times
10^{-24}~{\rm s}~.
\eeq
During this time it exerts a force due to the string tension $k = 0.18$
GeV$^2$ and hence imparts a momentum
\beq
\Delta p_{z} = k t = \frac{(0.18~{\rm GeV}^2)(5 \times 10^{-24}~{\rm s})}
{6.582 \times 10^{-25}~{\rm GeV}\cdot{\rm s}} \simeq 1.4~{\rm GeV}~
\eeq
to the $b$ quark, pulling it forward in the direction of the proton.  Since
the average $p_T$ of the $\Lambda_b$ in the CMS result is ${\cal O}(m_b)\simeq
5$ GeV (see Fig.\ \ref{fig:ptdist}), this should be a non-negligible effect.
Such ``string drag'' phenomena are taken into account in recent Monte Carlo
approaches \cite{Skands:2012mm}.

To compare with a result of Ref.\ \cite{Norrbin:2000jy}, we note that with
$y \equiv - \ln \tan (\theta/2)$, $dy/d\theta = - \cosh y$ which is $-1$
at $y=0$.  Here $\theta$ denotes the polar angle of the $b$ quark.  If $p_T$
is its transverse momentum, we have $\Delta \theta \simeq - \Delta p_z/p_T$
or, at $y=0$, $\Delta y = - \Delta \theta \simeq 1.4~{\rm GeV}/p_T$.  This
is approximately of the form found in Ref.\ \cite{Norrbin:2000jy}, but about
three times as large.  As the result is expressed in terms of boost-invariant
quantities, it is now valid for any $y$.

\section{Other distinguishing measurements}

\subsection{Transverse momentum and $|y|$}

We have mentioned that the string-drag mechanism leads to an effect $\Delta y
= -1.4$ GeV/$p_T$.  It is harder to separate the $|y|$ and $p_T$ dependences
of the quasi-diffractive excitation model.  One may think of the mechanism
of Fig.\ \ref{fig:diffrac} as the effect of diffractive excitation of many
$B^+ \Lambda_b$ resonances, in which case there are too many unknown variables
to permit quantitative estimates.  The difficulty of the problem is not
unlike that encountered in interpreting fixed-target hyperon production
(e.g., \cite{Abouzaid:2006ku} and references therein).  Nonetheless, one can
anticipate that the importance of quasi-diffractive excitation should
increase with decreasing $p_T$ and increasing $|y|$.  One may be able to
gauge its importance by looking for $B^+$--$\Lambda_b$ correlations, as
suggested by the picture of Fig.\ \ref{fig:diffrac}.

\subsection{$\Lambda_b$ polarization}

The fixed-target study of hyperons mentioned above \cite{Abouzaid:2006ku} and
earlier investigations turned up unexpectedly large transverse polarizations
without a clearly understood pattern.  In 500 GeV/$c$ $\pi^- N$ collisions,
$\Lambda_c$ polarization is found to become increasingly negative with
increasing $p_T$ \cite{Aitala:1999}.  A hybrid perturbative QCD model with
polarization transfer from $c$ to $\Lambda_c$ can account for this effect
\cite{Goldstein:1999}.  In contrast, no $\Lambda_b$ polarization
has been seen by any of the three LHC experiments.  CMS
\cite{Rikova:2013} finds $P(\Lambda_b) = 0.03 \pm 0.09 \pm 0.03$ and
$P(\bar \Lambda_b) = 0.02 \pm 0.08 \pm 0.05$; ATLAS \cite{Aad:2014iba} finds
both $P(\Lambda_b)$ and $P(\bar \Lambda_b)$ consistent with zero; and LHCb
\cite{Aaij:2013oxa} finds the polarizations of $\Lambda_b$ and $\bar \Lambda_b$
consistent with each other, giving an average of $0.06 \pm 0.07 \pm 0.02$.

One feature of $\Lambda_b$ polarization is that in the constituent-quark
picture, the spin of the $\Lambda_b$ is carried entirely by the $b$ quark, as
the $u$ and $d$ quarks are coupled up to spin zero.  This correlation is
largely borne out by explicit QCD calculations \cite{Mannel:1992,Falk:1993}.
Standard estimates of $\Lambda_b$ polarization at the LHC fall in the
10--20\% range \cite{Hiller:2007,Ajaltouni:2005}.

A relatively recent discussion of the induction of spin-spin forces by
exchange of a QCD string has been given in Ref.\ \cite{Vyas:2008}.  An
interesting feature, which unfortunately prevents a quantitative conclusion,
is that the effect behaves as the fourth power of the string thickness, an
unknown quantity.

\section{Conclusions}

Mechanisms have been described which favor forward production of heavy baryons
in $b$ quark fragmentation. These include quasi-diffractive processes, in which
a proton dissociates into a heavy baryon and a meson containing a $\bar b$, and
a string-drag effect \cite{Norrbin:2000jy,Andersson:1983ia} investigated
in the context of top quark production \cite{Rosner:2012pi}.  While found
to be unimportant in generating any forward-backward asymmetry at the Tevatron
for top production, the latter mechanism is seen to have greater effect
in generating an asymmetry in $\Lambda_b$ and $\bar \Lambda_b$ production.
Such an asymmetry is suggested in the highest-$y$ bin studied by the CMS
Collaboration \cite{Chatrchyan:2012xg}, where the cross section for
$\bar \Lambda_b$ production is about 2/3 that for $\Lambda_b$ production.

It would be extremely interesting to study these effects at ATLAS, LHCb, and the
Fermilab Tevatron, comparing them with available Monte Carlo predictions.  In
the latter, the quasi-diffractive and string-drag processes should generate a
leading baryon effect, in which the $\Lambda_b$ and $\bar \Lambda_b$ tend to
follow the direction of the proton and antiproton, respectively.  Such an
asymmetry is immune to systematic differences in detection efficiencies for
particles and antiparticles \cite{Lewis}, lending unique urgency to such
studies at the Tevatron.

\section*{Acknowledgments}

I thank Samim Erhan, Jonathan Lewis, Patrick Koppenburg, Patrick Lukens,
Brian Meadows, and Sheldon Stone for helpful advice.  This work was supported
in part by the U. S. Department of Energy under Grant No.\ DE-FG02-13ER41958.

\end{document}